\documentclass[letterpaper, 10 pt, conference]{ieeeconf}
\IEEEoverridecommandlockouts    
\usepackage{graphics}
\usepackage{epsfig}
\usepackage{times}
\usepackage{amsmath}
\usepackage{amssymb}

\makeatletter
\let\MYcaption\@makecaption
\makeatother

\usepackage[font=footnotesize]{subcaption}

\makeatletter
\let\@makecaption\MYcaption
\makeatother

\usepackage{mathtools}
\usepackage{color}
\usepackage{cite}
\usepackage{comment}
\usepackage{balance}
\usepackage[table]{xcolor}
\usepackage{multirow}
\usepackage{hyperref}
\hypersetup{
    colorlinks=true,
    linkcolor=black,
    citecolor=green,
    urlcolor=blue,
}
\usepackage{blindtext}
\usepackage{bm}
\usepackage{algorithm}
\usepackage{algpseudocode}
\usepackage[official]{eurosym}
\usepackage{multirow}
\usepackage{booktabs}

\algrenewcommand\algorithmicindent{1.0em}
\algnewcommand\algorithmicswitch{\textbf{switch}}
\algnewcommand\algorithmiccase{\textbf{case}}
\algnewcommand\algorithmicassert{\texttt{assert}}
\algnewcommand\Assert[1]{\State \algorithmicassert(#1)}%
\algdef{SE}[SWITCH]{Switch}{EndSwitch}[1]{\algorithmicswitch\ #1\ \algorithmicdo}{\algorithmicend\ \algorithmicswitch}%
\algdef{SE}[CASE]{Case}{EndCase}[1]{\algorithmiccase\ #1}{\algorithmicend\ \algorithmiccase}%
\algtext*{EndSwitch}%
\algtext*{EndCase}%

\makeatletter
\newcommand{\algmargin}{\the\ALG@thistlm}
\makeatother
\newlength{\forwidth}
\algdef{SE}[parFOR]{parFor}{EndparFor}[1]
  {\parbox[t]{\dimexpr\linewidth-\algmargin}{%
     \hangindent\forwidth\strut\algorithmicfor\ #1\ \algorithmicdo\strut}}{\algorithmicend\ \algorithmicfor}%
\newlength{\forif}
\algnewcommand{\parState}[1]{\State%
  \parbox[t]{\dimexpr\linewidth-\algmargin}{\strut #1\strut}}

\title{\LARGE \bf
Towards Safe Autonomous Intersection Management: \\
Temporal Logic-based Safety Filters for Vehicle Coordination
}

\author{Kaj Munhoz Arfvidsson, Frank J. Jiang, Karl H. Johansson, Jonas Mårtensson%
\thanks{
    This work was partially supported by the Wallenberg Artificial Intelligence, Autonomous Systems, and Software Program (WASP) funded by the Knut and Alice Wallenberg Foundation. It was also partially supported by the Swedish Research Council, Swedish Research Council Distinguished Professor Grant 2017-01078, the Knut and Alice Wallenberg Foundation Wallenberg Scholar Grant, and the Swedish Innovation agency (Vinnova), under grant 2021-02555 Future 5G Ride, within the Strategic Vehicle Research and Innovation program (FFI).}
\thanks{
    All authors are with the Division of Decision and Control Systems, EECS, KTH Royal Institute of Technology, Malvinas v{\"a}g 10, 10044 Stockholm, Sweden {\tt\small \{kajarf, frankji, kallej, jonas1\}@kth.se}. They are also affiliated with the Integrated Transport Research Lab and Digital Futures.}%
}

\begin{document}

\maketitle
\thispagestyle{empty}
\pagestyle{empty}

\begin{abstract}
In this paper, we introduce a temporal logic-based safety filter for Autonomous Intersection Management (AIM), an emerging infrastructure technology for connected vehicles to coordinate traffic flow through intersections. Despite substantial work on AIM systems, the balance between intersection safety and efficiency persists as a significant challenge. Building on recent developments in formal methods that now have become computationally feasible for AIM applications, we introduce an approach that starts with a temporal logic specification for the intersection and then uses reachability analysis to compute safe time-state corridors for the connected vehicles that pass through the intersection. By analyzing these corridors, in contrast to single trajectories, we can make explicit design decisions regarding safety-efficiency trade-offs while taking each vehicle's decision uncertainty into account. Additionally, we compute safe driving limits to ensure that vehicles remain within their designated safe corridors. Combining these elements, we develop a service that provides safety filters for AIM coordination of connected vehicles. We evaluate the practical feasibility of our safety framework using a simulated 4-way intersection, showing that our approach performs in real-time for multiple scenarios.
\end{abstract}

\newcommand{\BRS}{\text{BRS}}
\newcommand{\BRSmax}{\BRS^{max}}
\newcommand{\BRSmin}{\BRS^{min}}
\newcommand{\BRT}{\text{BRT}}
\newcommand{\BRTmaxany}{\BRT^{max}_{any}}
\newcommand{\BRTmaxall}{\BRT^{max}_{all}}
\newcommand{\BRTminany}{\BRT^{min}_{any}}
\newcommand{\BRTminall}{\BRT^{min}_{all}}

\newcommand{\ltltrue}{\textit{true}}
\newcommand{\ltlfalse}{\textit{false}}
\newcommand{\ltlnot}{\neg}
\newcommand{\ltlor}{\lor}
\newcommand{\Ltlor}{\bigvee}
\newcommand{\ltland}{\land}
\newcommand{\Ltland}{\bigwedge}
\newcommand{\ltlimply}{\rightarrow}
\newcommand{\ltlnext}{\bigcirc}
\newcommand{\ltluntil}{\,\mathsf{U}\,}
\newcommand{\ltlalways}{\square}
\newcommand{\ltleventually}{\lozenge}
\newcommand{\ltlsatisfy}{\models}

\section{Introduction}\label{sec:intro}

Over the last decade, a new generation of Intelligent Transportation Systems (ITS) has begun to emerge, driven by advances in computing and networking \cite{arthurs_taxonomy_2022}. Vehicles are expected to communicate with each other, infrastructure, and the cloud to improve safety, efficiency, sustainability, and passenger comfort \cite{soto_survey_2022}. Furthermore, with the rise of edge- and cloud-services, significant effort has been directed into facilitating these objectives using off-board intelligence in the local infrastructure \cite{arthurs_taxonomy_2022,gong_edge_2023}. Autonomous Intersection Management (AIM) has, in particular, received considerable attention due to the complex challenges that arise at intersections and the limited time that vehicles spend there \cite{chen_cooperative_2016}. 
Intelligent intersections have shown that they can, even in mixed-traffic, collaborate with partially automated vehicles to act as a safety filter, alerting or possibly overriding the human driver if deemed unsafe \cite{ahn_semi-autonomous_2016, altche_least_2016}.
Notably, some researchers state that safety should never be compromised to achieve efficiency~\cite{dresner_multiagent_2008,chamideh_safe_2023}. However, although many proposed systems claim such improvements, the trade-off between safety and efficiency still remains an open challenge and an important research direction \cite{namazi_intelligent_2019}.

\begin{figure}[t]
    \centering
    \includegraphics[width=\linewidth]{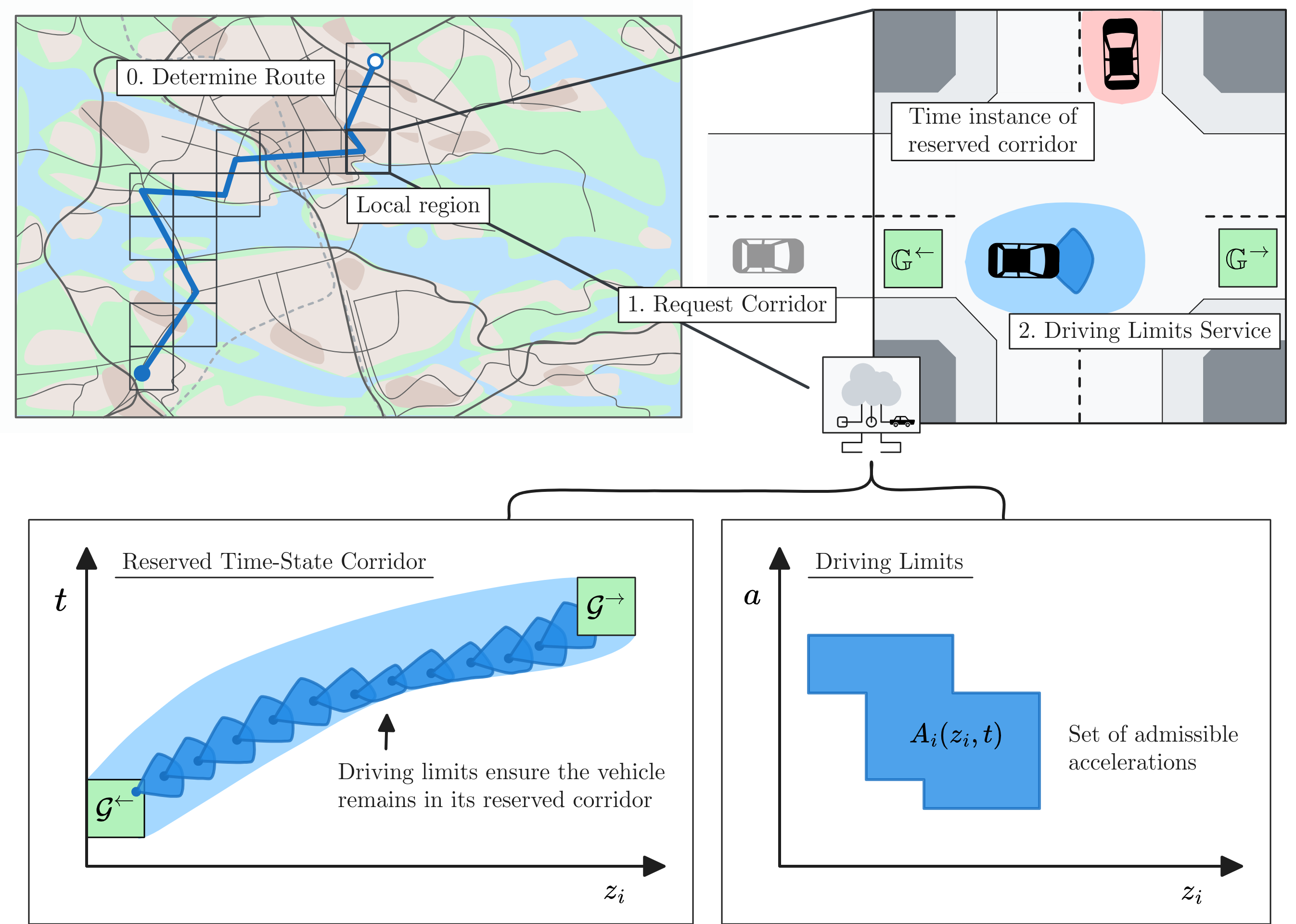}
    \caption{Shown is the conceptual design of our traffic management system. While acting between high-level city-scale route planning and on-board control, this system performs central planning and coordination between vehicles for local regions while accounting for their safe operational limits.}
    \label{fig:ltms}
\end{figure}

Proving that behavior is safe poses major challenges \cite{colombo_efficient_2012}. To this end, formal methods have been used to ensure safe collision avoidance \cite{hafner_cooperative_2013}, perform automata-based verification of collision zones \cite{saraoglu_designing_2022}, and construct specification-compliant driving corridors \cite{irani_liu_specification-compliant_2023}. Though, it is only now that such approaches have become computationally tractable for real-time AIM applications, particularly for high-dimensional, multi-agent scenarios that use nonlinear vehicle models. Building on this opportunity, researchers recently developed a method that addresses the safety challenge of intersections
\cite{munhoz_ensuring_2024}. Motivated in part by the difficulty to extend and adapt the control architectures of traditional traffic management systems, their method uses reachability analysis to compute all safe trajectories of a vehicle. From these, it is possible to freely pick which one to follow, making the approach suitable to adapt or extend to other control architectures.
 
In this work, we aim to extend~\cite{munhoz_ensuring_2024} to help develop provably safe AIM systems for intelligent intersections. In this extension, we further expand on the practicalities of the safety framework to enable implementations and integration with AIM systems. Notably, a significant contribution over~\cite{munhoz_ensuring_2024} is the design and development of a driving limits service that connected vehicles can use to receive safe acceleration bounds that guarantee 
safe behavior
and contribute to improving the intersection's overall efficiency. In summary, this paper's contributions are three-fold:
\begin{enumerate}
    \item we build upon the previous safety framework in \cite{munhoz_ensuring_2024} 
    so that AIMs can take conscious design decisions about the trade-off between safety and efficiency,
    \item we design 
    a driving limits service
    that further facilitate the development of safe AIMs,
    \item we evaluate the practical feasibility of the safety framework in multiple scenarios of a 4-way intersection.
\end{enumerate}
To achieve strong safety guarantees, noted as a requirement by \cite{dresner_multiagent_2008,chamideh_safe_2023}, our approach formalizes a safety specification using temporal logic statements.
Using temporal logic trees, we verify the specification with Hamilton-Jacobi (HJ) reachability analysis, also enabling efficient, real-time computation of safe acceleration sets~\cite{jiang_guaranteed_2024}.
Although the computational complexity of these methods have previously been impractical, preliminary indications now show that computation time is fast enough for real systems.
By providing these features, which are both important and practically necessary for intelligent intersections, we help accelerate the development of AIM and vehicle automation generally.

\section{Preliminaries}\label{sec:prelim}

Much of the preliminary material is derived from~\cite{munhoz_ensuring_2024}, thus, for brevity, we have excluded several technical details in this preliminary section and focus on presenting the differences and key concepts that are needed for this work.

\subsection{Intersection Model}
In this work, we use a similar model to the one used in~\cite{munhoz_ensuring_2024}. That is, for an intersection with $N$ vehicles, denote the collective states and control inputs of all vehicles with $z = [z_1, z_2, \ldots, z_N]^\top$ and $u = [u_1, u_2, \ldots, u_N]^\top$. We then denote full intersection dynamics with the following:
\begin{equation}\label{eq:multi_vehicle}
    \dot{z} = f(z, u) + g(z)u.
\end{equation}
To ensure our method takes into account the common nonlinear behaviors vehicles exhibit, we model each individual vehicle with a four state nonlinear bicycle model. Specifically, for a single vehicle $i$, let $z_i = [x_i, y_i, \theta_i, v_i]^\top$ be the state, where $x_i$, $y_i$, $\theta_i$, and $v_i$ are the vehicle's x-position, y-position, heading angle, and velocity, respectively. Then, let $u_i = [\delta_i, a_i]^\top$ be the input, where $\delta$ and $a$ are the steering and acceleration inputs of the vehicle. While similar to~\cite{munhoz_ensuring_2024}, in this work, we reduce the state dimension by one to significantly decrease computational complexity. Without sacrificing any nonlinear dynamics,
\begin{equation}
    f_i(z_i, u_i) =
    \left[\begin{matrix}
        v_i \cos\theta_i \\
        v_i \sin\theta_i \\
        \frac{v_i \tan \delta_i}{L_i}\\
        0
    \end{matrix}\right], \ 
    g_i(z_i) = 
    \left[\begin{matrix}
        0 & 0\\
        0 & 0\\
        0 & 0\\
        0 & 1
    \end{matrix}\right],
    \nonumber
\end{equation}
where $L_i$ is the wheel-base length of the vehicle. This reduced model is only control-affine with respect to the acceleration input, which is sufficient for this work.

\subsection{Intersection Verification}

In this work, we build on the verification pipeline formulated in~\cite{munhoz_ensuring_2024}. This pipeline is flexible and general enough to incorporate many of the detailed requirements one may have for AIM. It is summarized by the following:
\begin{enumerate}
    \item \textbf{Intersection Behavior Specification}: desired and required behavior is specified in the well-known, formal system called linear temporal logic (LTL). Using LTL, there exist methods to automatically verify that vehicles can conform to the desired and required behaviors of the intersection.
    \item \textbf{Temporal Logic Tree Construction}: once the intersection behavior specification is determined, we apply a computational verification approach using temporal logic trees. This involves constructing a temporal logic tree with reachability analysis to ensure that trajectories exist which satisfy the behavior specification.
    \item \textbf{Formal Verification Check}: after constructing the temporal logic tree, we verify the feasibility of all vehicles safely passing through the intersection by examining the state set in the root node. The volume of this state set corresponds to whether there exist trajectories for each vehicle that fully and safely satisfy the specified behaviors and requirements.
\end{enumerate}
The use of temporal logic specifications for intersections allow us to easily define and modify new requirements for an intersection based on changes in desired behavior or updates in legal requirements. Due to these benefits, there are recent works leveraging temporal logic specifications for encoding and formally verifying requirements in ITS, e.g.~\cite{maierhofer_formalization_2020, saraoglu_designing_2022}. 

To construct temporal logic trees, we utilize HJ reachability analysis to compute the backward reachable tube,
\[\begin{split}
    \mathcal R_B(\mathcal G; \mathcal C) = \{ 
        (z, t) \; | \;
            & \exists u(\cdot) \in U, \: \\
            & \exists (z_f, t_f) \in \mathcal{G}, \: \zeta(t_f; z, t, u(\cdot)) = z_f, \\
            & \forall \tau \in [t, t_f), \:
            \zeta(\tau; z, t, u(\cdot)) \in \Omega_{\mathcal C}(\tau)
    \},
\end{split}\]
where $\mathcal G$ and $\mathcal C$ are goal and constraint time-state sets corresponding to particular requirements encoded in the intersection specification. Using the notation from \cite{munhoz_ensuring_2024}, $\Omega_\mathcal{C}(\cdot)$ is a map from time to its corresponding state set. $u(\cdot)$ and $\zeta(\cdot)$ denote a control policy and the resultant trajectory of that control policy. Intuitively, $\mathcal R_B(\cdot)$ is a set that includes all time-states where vehicles in the intersection can satisfy a particular requirement. For more details about the full definition of this backward reachable tube, we refer readers to~\cite{munhoz_ensuring_2024}. Due to the simplification presented in~\cite[Section III.D]{munhoz_ensuring_2024}, we do not explicitly define the commonly computed robust controlled invariant set here. When we refer to these conceptually, we denote them with $\mathcal{RCI}(\cdot)$.

\subsection{Forward Reachability Analysis}
While backward reachability analysis is enough to verify the safety and feasibility of all vehicles in the intersection, in this work we explore the use of forward reachability analysis to improve throughput and efficiency. Specifically, instead of allowing for vehicles to enter and exit freely, we will use estimated time windows of entrance and exit to narrow their planned occupancy. We iteratively perform additional reachability analyses to refine this occupancy with added time constraints. For exit constraints, we can reuse the computation for $\mathcal{R}_B(\cdot)$. However, for entrance constraints, we must follow trajectories of the system forwards in time. Explicitly, we compute the forward reachable tube,
    \[\begin{split}
    \mathcal R_F(\mathcal I; \mathcal C) = \{ 
        (z, t) \; | \;
            & \exists u(\cdot) \in U, \\
            & \exists (z_0, t_0) \in \mathcal{I}, \: \zeta(t; z_0, t_0, u(\cdot)) = z, \\
            & \forall \tau \in [t_0, t], \:
            \zeta(\tau; z_0, t_0, u(\cdot)) \in \Omega_{\mathcal C}(\tau)
    \},
    \end{split}\]
where $\mathcal I$ and $\mathcal C$ are specific initial and constraint time-state sets corresponding to a particular vehicle's entrance time window and encoded intersection constraints.

\begin{figure*}[t]
    \centering
    \begin{subfigure}{0.24\textwidth}
        \centering
        \includegraphics[width=\textwidth]{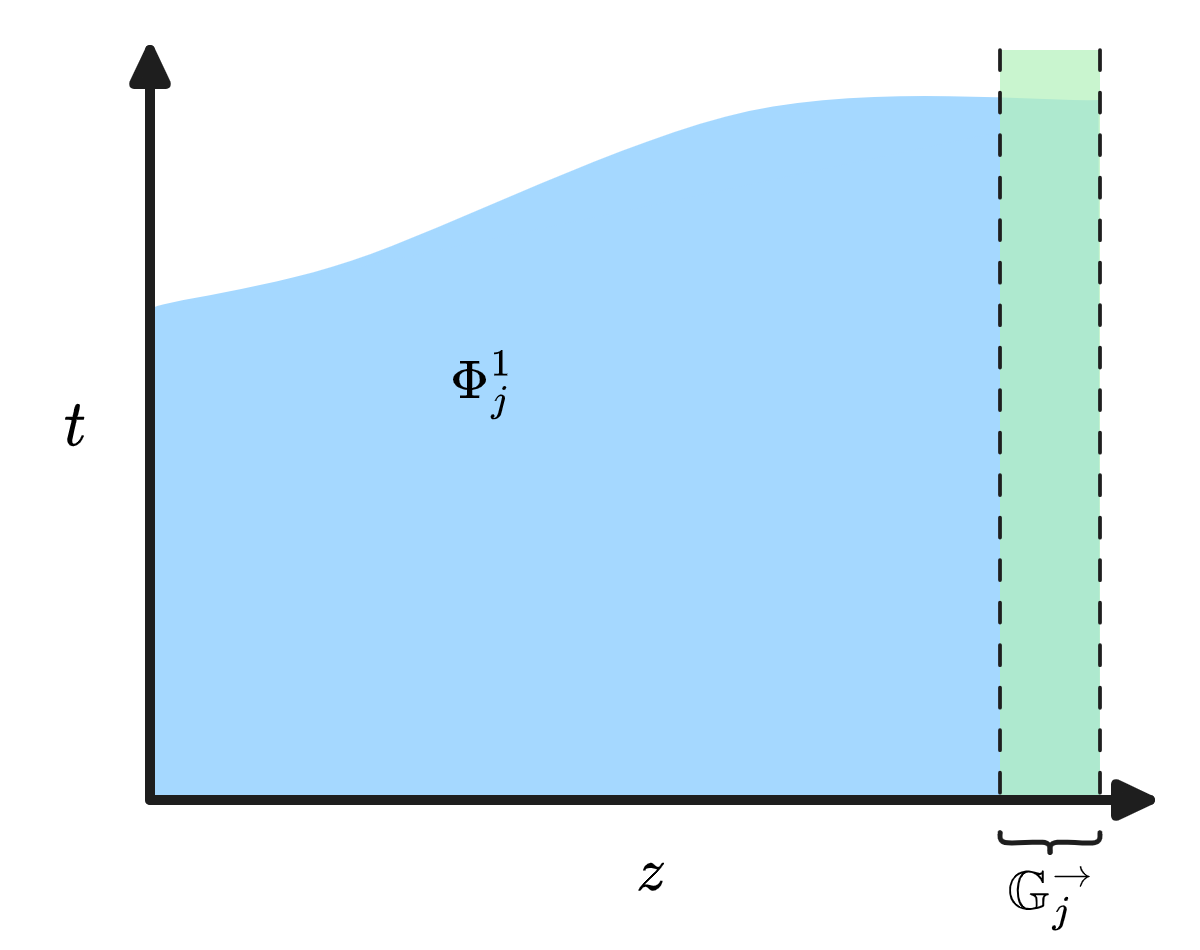}
        \caption{\textit{Pass 1}.}
        \label{fig:pass1}
    \end{subfigure}\hfill
    \begin{subfigure}{0.24\textwidth}
        \centering
        \includegraphics[width=\textwidth]{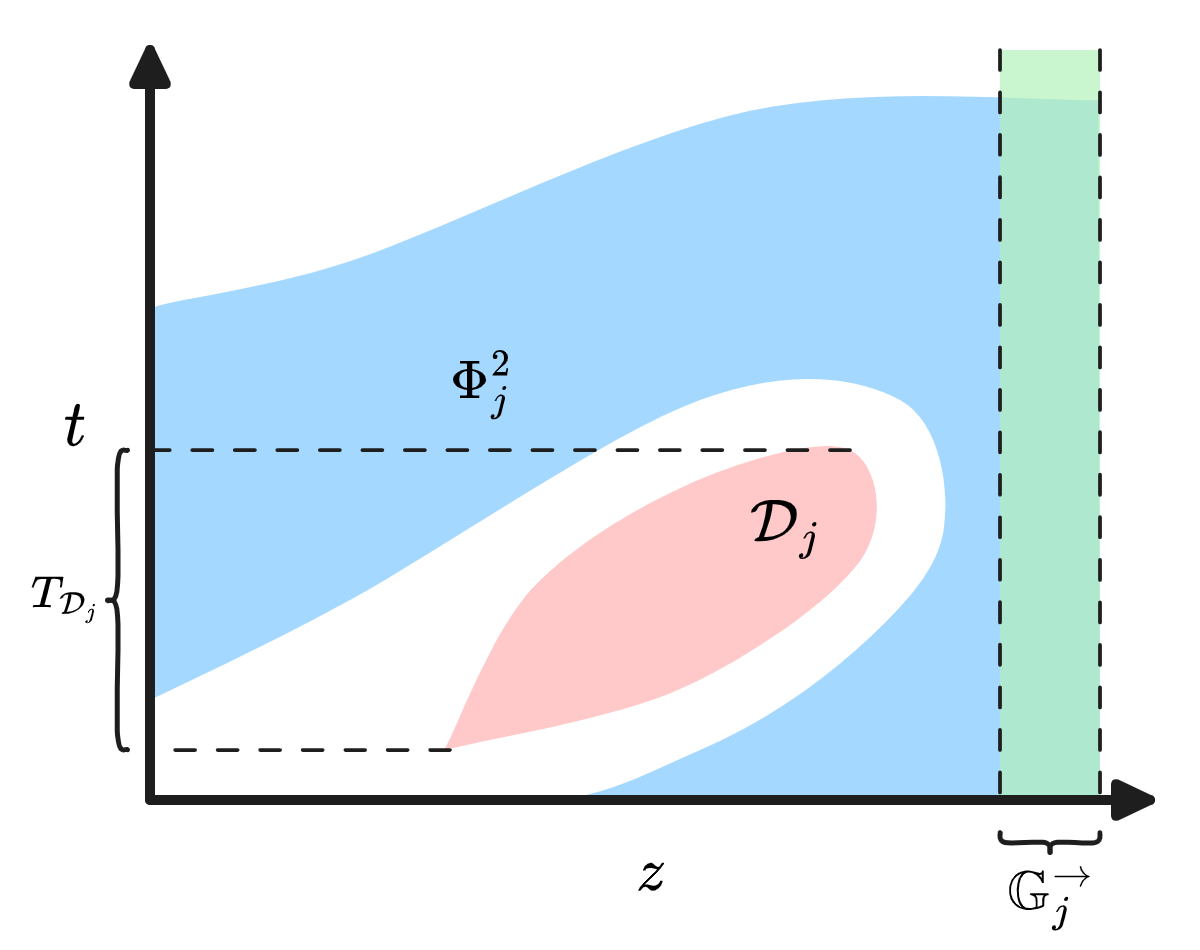}
        \caption{\textit{Pass 2}.}
        \label{fig:pass2}
    \end{subfigure}\hfill
    \begin{subfigure}{0.24\textwidth}
        \centering
        \includegraphics[width=\textwidth]{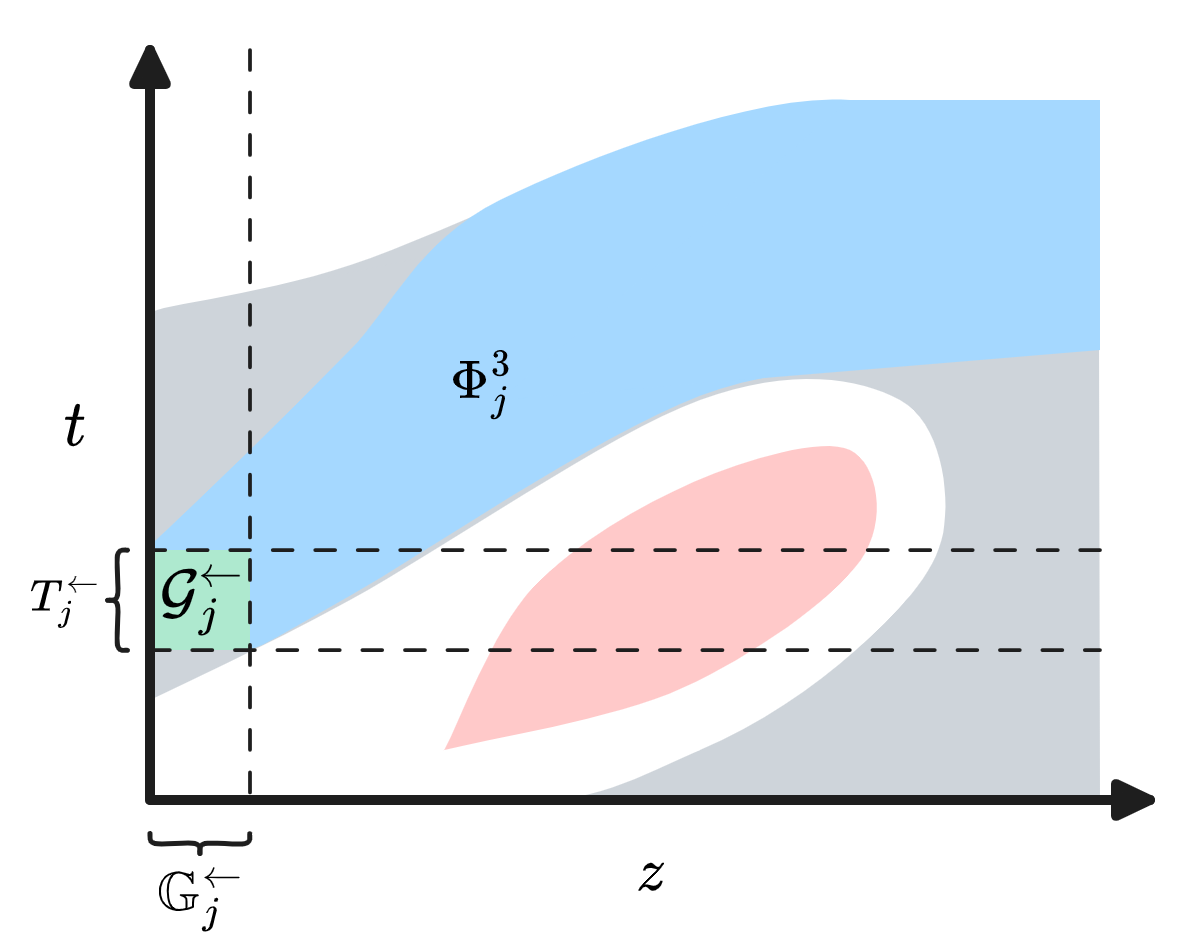}
        \caption{\textit{Pass 3}.}
        \label{fig:pass3}
    \end{subfigure}\hfill
    \begin{subfigure}{0.24\textwidth}
        \centering
        \includegraphics[width=\textwidth]{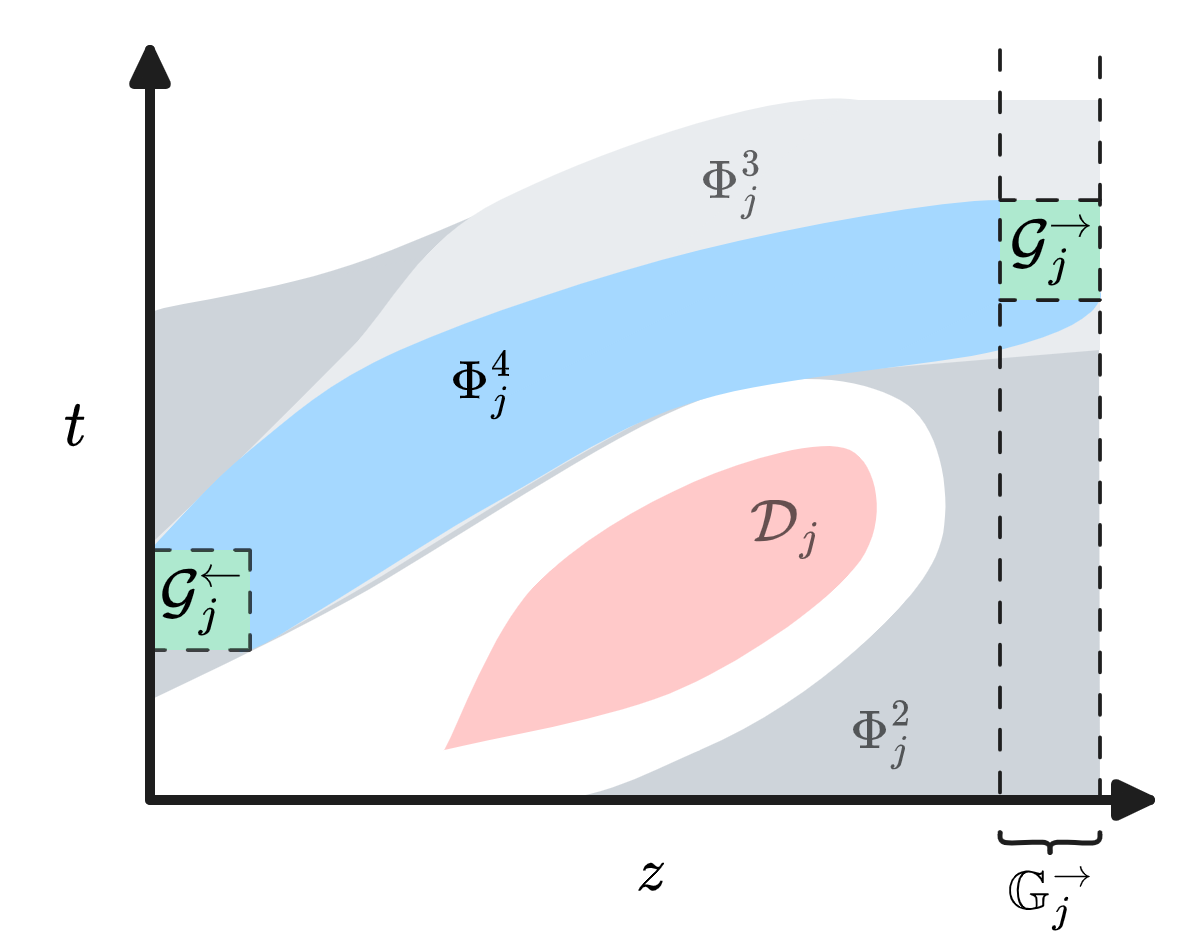}
        \caption{\textit{Pass 4}.}
        \label{fig:pass4}
    \end{subfigure}
    \caption{
        Conceptual illustration of the four passes in our analysis.
        (a) shows $\Phi^1_j$, the result of \textit{Pass 1}, which ensures conformity to invariant constraints, such as road geometries or static obstacles. This is illustrated by its shape on the top.
        (b) shows $\Phi^2_j$ when non-invariant constraints are added in the form of $\mathcal D_j$.
        (c) shows $\Phi^3_j$ in the first step of refining the trajectories. After this pass, all remaining trajectories that are embedded in $\Phi^3_j$ satisfy the entry conditions $\mathcal G^\leftarrow_j$.
        (d) shows $\Phi^4_j$ after it has similarly pruned trajectories even further, now also considering exit conditions $\mathcal G^\rightarrow_j$.
    }
    \label{fig:passes}
\end{figure*}

\subsection{Safe Driving Limits}

One key advantage of using HJ reachability analysis for the intersection verification is the ability to use constructed temporal logic trees to efficiently compute input saturation points that guarantee safety and satisfaction of the intersection specification in real-time~\cite{jiang_guaranteed_2024}. In other words, we are able to compute driving limits that vehicles must respect to meet the intersection requirements. As long as driving inputs are within these saturation points, safety is guaranteed, providing vehicles with the maximum freedom to operate. Specifically, for model~\eqref{eq:multi_vehicle}, this means that after specifying and verifying the desired intersection behavior, we can compute the minimum and maximum acceleration for each vehicle passing through the intersection.
Namely, for vehicle state $z_j$ and time point $t$, let $A_j(z_j, t)$ be the set of accelerations that vehicle $j$ can implement to be guaranteed to satisfy the intersection specification.

\section{Motivation}\label{sec:mot}

To contextualize our work, consider a vehicle following some global route through a city as illustrated in Fig.~\ref{fig:ltms}. Like with other traffic infrastructure, we imagine that intelligent services are provided at local regions throughout the city. Specifically, we assume that the intersection in Fig.~\ref{fig:ltms} implements reservation-based AIM for the local traffic \cite{chen_cooperative_2016,namazi_intelligent_2019}. When approaching, the vehicle requests a safe corridor through the region by providing information about when and where it will enter and exit the intersection. While the vehicle is inside the intersection, it is required to stay within the reserved corridor to ensure its own safety and that of others. As shown in Fig.~\ref{fig:ltms}, this is achieved using services that issue driving limits. Notably, the intelligent intersection coordinates these reservations and ensures the corridors are safe.

In our previous work \cite{munhoz_ensuring_2024}, we developed an approach to verify safety by analyzing collections of safe trajectories. Though this method could be used in AIM, if not given proper consideration, it could prove to be overly conservative. For example, consider reserving, in space and time, the entire collection of safe trajectories for each passing vehicle. Although the intersection would ensure safety of all managed vehicles, this approach would be overly conservative and block following vehicles. Specifically, the problem arises because many trajectories can satisfy a given safety specification but disregard overall efficiency. Imagine, for instance, a vehicle driving very slowly through the intersection. This behavior is sometimes necessary as it might be the only safe way through the area. However, when unnecessary, this behavior should not be allowed as it obstructs following vehicles and reduces intersection throughput. In essence, this is a trade-off relating to how permissive the intelligent intersection is designed to be. Throughout the rest of this work, we will refer to this as ``the problem of permissible planning.'' This paper explores how intelligent intersections can ensure safety, yet make conscious design decisions about the trade-off between safety and efficiency by addressing the problem of permissible planning.

\section{Ensuring Safety for AIM}\label{sec:method}

In this section, we describe our approach for ensuring safety for AIM. We first outline the high-level system design. Continuing from \cite{munhoz_ensuring_2024}, we then present how the system can sequentially ensure specification satisfaction for vehicles at the intersection. Inspired by commonly proposed AIM approaches, we then incorporate a coordination scheme that constructs provably safe corridors through the region based on entry and exit conditions. Finally, we demonstrate using a driving limits service how these corridors can provide safety filters for intelligent intersections.

\subsection{System Design}

For an approaching vehicle $j$, we suggest the following:
\begin{enumerate}
    \item The vehicle notifies the intelligent intersection
    that it will enter the region within time window $T^\leftarrow_j$ at state set $\mathbb G^\leftarrow_j$. Together with a requested exit $\mathbb G^\rightarrow_j$, this is necessarily communicated to pass through the region.
    \item Taking already scheduled vehicles into account, the safety analysis of \cite{munhoz_ensuring_2024} produces a safe time-state corridor for the approaching vehicle $j$.
    \item Now, to address the problem of permissible planning, the trajectories are further pruned such that only those satisfying the entry and exit conditions imposed by $T^\leftarrow_j$, $\mathbb G^\leftarrow_j$ and $\mathbb G^\rightarrow_j$ remain. Additionally, by following these trajectories, that start sometime in $T^\leftarrow_j$, we implicitly get an exit time window $T^\rightarrow_j$. 
    \item These results are then stored for as long as vehicle $j$ is in the region. Using them, the intelligent intersection computes safe control bounds. If followed, these ensure that vehicle $j$ stays within the reserved corridor.
\end{enumerate}

While we will not address inter-region management in this paper, we suggest that our approach provides the necessary interfaces for composition. That is, when going between regions $A$ and $B$ that overlap in $\mathbb G^{AB}$, then the handover could be done in $T^{AB} \times \mathbb G^{AB}$, where $T^{AB}$ is the time window for exiting region $A$ and entering region $B$. Since entry time drives exit time, $T^{AB} \subseteq T^{A\rightarrow}$. However, to discuss inter-regional aspects, we must first formalize our approach for a single region.

\subsection{Verification of Safety Specification}

The approach in \cite{munhoz_ensuring_2024} is largely based on formalizing the sequential path planning (SPP) method of \cite{bansal_provably_2021} as a temporal logic specification. We summarize the key points here for the reader's convenience.

SPP is a priority-based method that treats each vehicle sequentially according to some priority order. For example, priority could be given as per first-come-first-served (FCFS) principles. While planning for vehicle $j$, since trajectories of higher priority vehicles $i<j$ are already known, they can be considered as deterministic obstacles. This is captured by the LTL formula
\begin{equation}\label{eq:ltl}
	\varphi_j = \ltleventually g_j \ltland \ltlalways c_j \ltland \ltlalways \ltlnot \Ltlor_{i<j} d_{j,i},
\end{equation}
where $\ltleventually g_j$ and $\ltlalways c_j$ respectively requires that vehicle $j$ eventually reaches its goal and that it always obeys traffic rules. Specifically, propositions $g_j$ and $c_j$ correspond to certain $\mathcal{G}_j$ and $\mathcal{C}_j$ that, respectively, encode goals and constraints in time-state sets. The final term relates to collision avoidance of higher priority vehicles that are similarly encoded in danger sets $\mathcal{D}_{j,i}$, for all vehicles $i<j$. Like reservation-based approaches for AIM, these danger sets contain the reserved corridors for higher priority vehicles. Then, due to the sequential evaluation, it is sufficient to only check that vehicle $j$ does not enter any of the danger sets. This recursively ensures that no vehicles will collide.

To analyze the specification \eqref{eq:ltl}, we construct a temporal logic tree by replacing propositions with corresponding time-state sets, logical operators with set operations and temporal operators with reachability analyses. For instance, $c_j$ relates as mentioned to $\mathcal{C}_j$ and $\Ltlor_{i<j} d_{j,i}$ correspond to $\bigcup_{i<j} \mathcal D_{j,i}$. The temporal operators, ``eventually'' ($\ltleventually$) and ``always'' ($\ltlalways$) typically correspond to computing $\mathcal{R}_B(\cdot)$ and $\mathcal{RCI}(\cdot)$, respectively. However, in this case, we can evaluate the expression $\ltleventually g_j \ltland \ltlalways c_j$ with only one reachability analysis, namely $\mathcal{R}_B(\mathcal{G}_j; \mathcal{C}_j)$. We refer to~\cite[Section III.D]{munhoz_ensuring_2024} for more information on this. When incorporating the danger sets $\mathcal{D}_j = \bigcup_{i<j}\mathcal{D}_{j,i}$ and evaluating the full safety specification~\eqref{eq:ltl}, the temporal logic tree similarly results in the computation of the time-state set
\begin{equation}\label{eq:reach}
    \mathcal{R}_B(\mathcal{G}_j ; \mathcal{C}_j \cap \mathcal{D}^C_j).
\end{equation}
If inside this set, vehicle $j$ is able to satisfy~\eqref{eq:ltl}, including the objectives of reaching its goal location while staying safe in terms of collision avoidance and obeying traffic rules.

While we formally wish to compute~\eqref{eq:reach}, higher priority vehicles will not necessarily interact with vehicle $j$. That is, $\mathcal D_j$ will often only impose constraints during a certain time window $T_{\mathcal D_j}$. Consequently, at times $\mathcal D_j = \emptyset$ which in turn simplifies~\eqref{eq:reach} to $\mathcal{R}_B(\mathcal{G}_j; \mathcal{C}_j)$. Moreover, since lane geometries, traffic rules and other state constraints that constitute $\mathcal C_j$ are often static, it is possible to precompute $\mathcal{R}_B(\mathcal{G}_j; \mathcal{C}_j)$. This leads to a potential performance gain where $\mathcal{R}_B(\mathcal{G}_j; \mathcal{C}_j)$ is precomputed offline, fetched from a lookup table at runtime and then updated with~\eqref{eq:reach} only for the relevant time window $T_{\mathcal D_j}$. Hence, the purpose of the update is to remove all trajectories that would enter $\mathcal D_j$ and collide with higher priority vehicles. As such, \cite{munhoz_ensuring_2024} introduces the ``Offline Pass'' and the ``Online Pass'' to denote the precomputation of $\Phi_j^1 = \mathcal{R}_B(\mathcal{G}_j; \mathcal{C}_j)$ and the subsequent online update using~\eqref{eq:reach} that produces $\Phi_j^2 \subseteq \Phi_j^1$. In this work, we will refer to these as \textit{Pass~1} and \textit{Pass~2}, respectively, and they are illustrated in Fig.~\ref{fig:pass1} and~\ref{fig:pass2}. For more details on the safety verification, we refer readers to \cite{munhoz_ensuring_2024}. 

\begin{figure*}[t]
    \centering
    \begin{subfigure}{0.32\textwidth}
        \centering
        \includegraphics[width=1.1\textwidth]{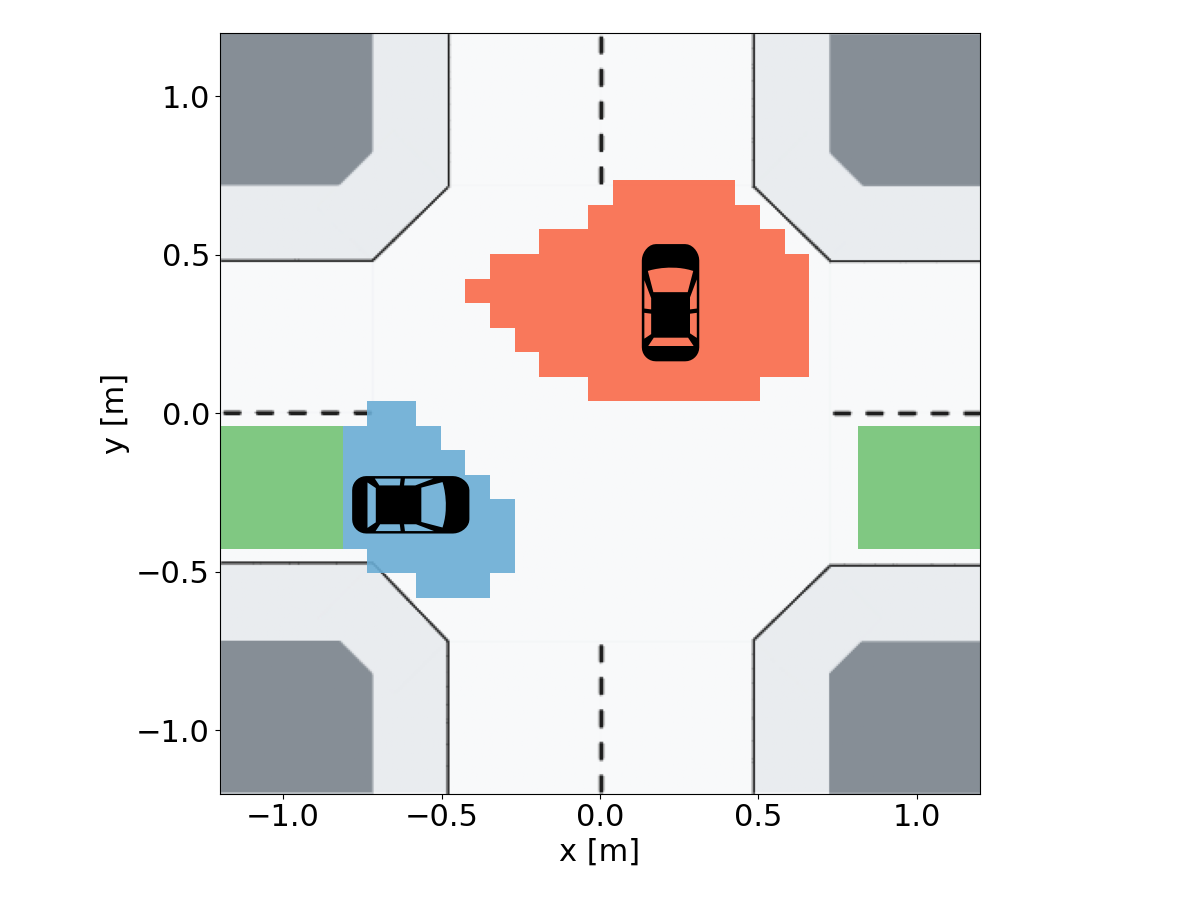}
        \caption{Seen from above, we see the safe corridors of both vehicles at a time instance. Additionally, entry and exit locations for the second vehicle are shown in green.}
        \label{fig:perp-xy}
    \end{subfigure}
    \hfill
    \begin{subfigure}{0.32\textwidth}
        \centering
        \includegraphics[width=1.1\textwidth]{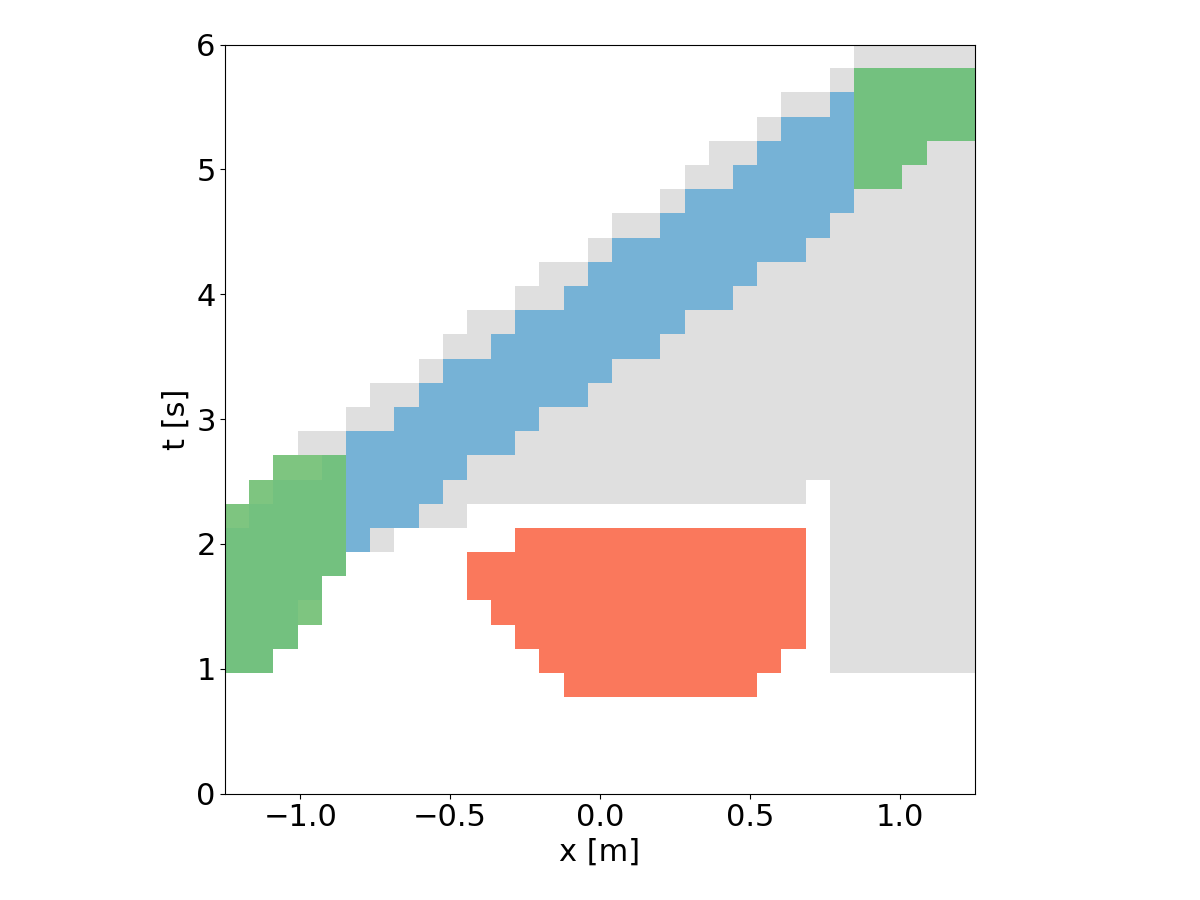}
        \caption{Slicing the intersection at $y=-0.3$, we view the progression of the sets in $x$ through time. We compare the result of \textit{Pass 4} (blue) with that of \textit{Pass 2} (gray).}
        \label{fig:perp-xt}
    \end{subfigure}
    \hfill
    \begin{subfigure}{0.32\textwidth}
        \centering
        \includegraphics[width=1.1\textwidth]{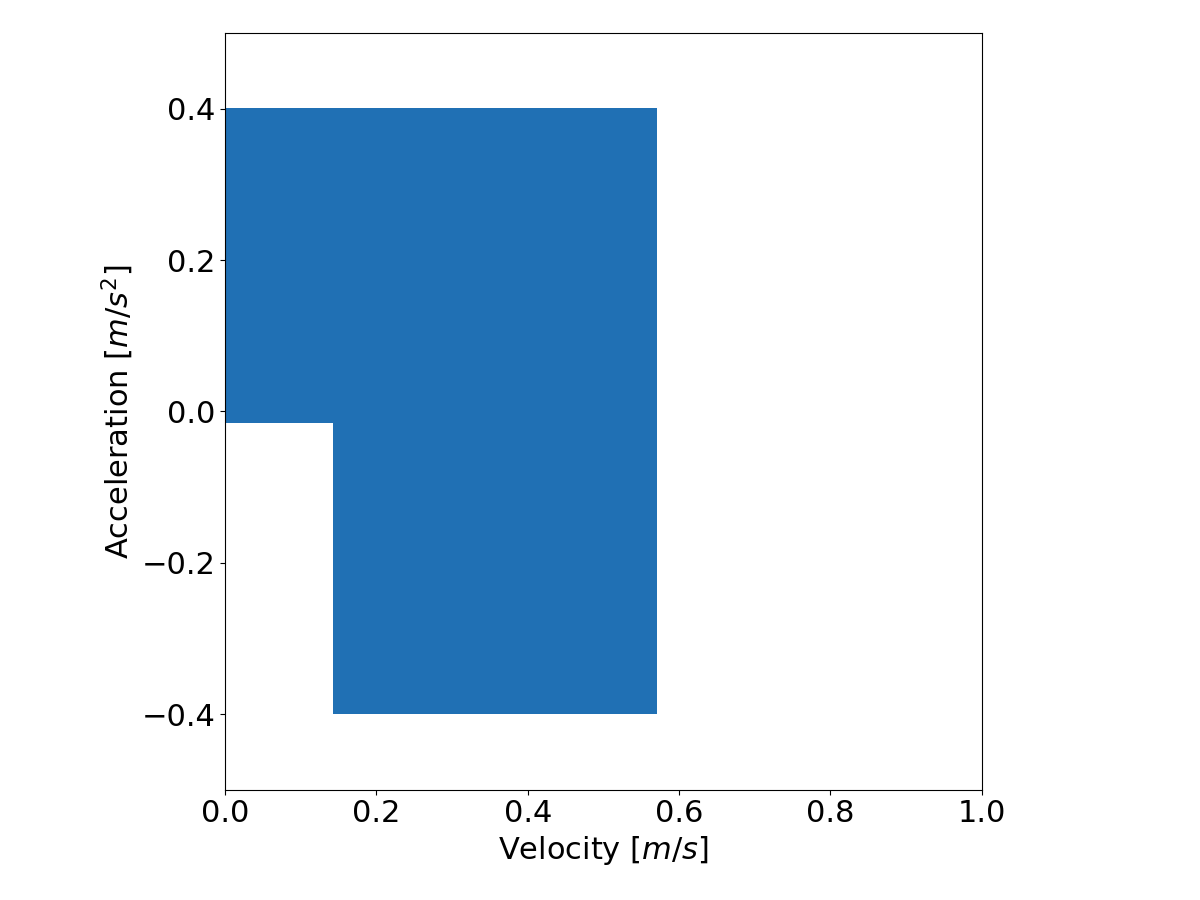}
        \caption{A set of admissible accelerations computed by the driving limits service for the second vehicle at $x_2=-0.75$, $y_2=-0.25$ and $\theta_2=0$.\\}
        \label{fig:lrcs}
    \end{subfigure}
    \caption{Two vehicles passing through the 4-way intersection perpendicular to each other. The first vehicle (red) going in the vertical direction, and the second vehicle (blue) going in the horizontal direction. Both vehicles stay within their indicated reserved time-state sets. In more detail, we show important aspects of the second vehicle's safety analyses.}
    \label{fig:perp}
\end{figure*}

\subsection{Refining Admissible Trajectories}

As mentioned in Section~\ref{sec:mot}, although the steps so far produce time-state sets that ensure safe behavior, ~\eqref{eq:ltl} also allows for behavior that negatively impacts throughput. This is a drawback of sequentially planning in general; we are not considerate of lower priority vehicles and the system as a whole. 
We will now introduce two important additions that refine the admissible trajectories of higher priority vehicles, \textit{Pass 3} and \textit{Pass 4}.

Given entry location $\mathbb G^\leftarrow_j$ and time window $T^\leftarrow_j$ for vehicle $j$, we want to find all trajectories that start from the entry time-state set $\mathcal G^\leftarrow_j = T^\leftarrow_j \times \mathbb G^\leftarrow_j$ and still satisfy the constraints imposed by~\eqref{eq:ltl}. The latter is true as long as the vehicle stays within $\Phi^2_j$. To achieve the former, we need to evaluate trajectories using forward reachability analysis. Specifically, \textit{Pass 3} computes 
\begin{equation*}    
    \Phi^3_j = \mathcal R_F(\mathcal G^\leftarrow_j; \Phi^2_j).
\end{equation*}
The resulting time-state set, illustrated in Fig.~\ref{fig:pass3}, starts at the entry and grows into $\Phi^2_j$. This highlights a common aspect of AIMs. Even though it is the vehicle that suggests entry conditions, there is a requirement that $\mathcal G^\leftarrow_j \subseteq \Phi^2_j$. If this is not the case, there exists no safe entry that both vehicle $j$ and the intelligent intersection can agree on. This could, for example, trigger a renegotiation depending on the control architectures of the intelligent intersection.

After \textit{Pass 3}, vehicle $j$ must enter the region at $\mathcal G^\leftarrow_j$ and exit at location $\mathbb G^\rightarrow_j$. However, $\Phi^3_j$ still allows the vehicle to exit at a number of different times. The last step in pruning the trajectories is done by selecting a time window $T^\rightarrow_j$, and consequently $\mathcal G^\rightarrow_j = T^\rightarrow_j \times \mathbb G^\rightarrow_j$, for exiting the region. However, deciding $T^\rightarrow_j$ can be done in different ways, either manually or automatically. In this work, we suggest finding the earliest possible time window of length $\Delta$. Using the time-state sets, this is easily done by finding the earliest time we can safely reach the exit:
\begin{equation*}\begin{split}
    \mathcal E &= \Phi^3_j \cap (T_{\Phi^2_j} \times \mathbb G^\rightarrow_j), \\
    t &= \inf \{ \tau \in T_{\mathcal E} \mid [\tau, \tau + \Delta] \subseteq T_\mathcal{E} \}. \\
\end{split}\end{equation*}
From this, we get the exit time window $T^\rightarrow_j = [t, t + \Delta]$.

With $\mathcal G^\rightarrow_j = T^\rightarrow_j \times \mathbb G^\rightarrow_j$ determined, we reach the final step of the refinement. Simply, in \textit{Pass 4} we once again perform another backward reachability analysis:
\begin{equation*}
    \Phi^4_j = \mathcal R_B(\mathcal G^\rightarrow_j, \Phi^3_j).
\end{equation*}
As with the entry, it is required that $\mathcal G^\rightarrow_j \subseteq \Phi^3_j$, meaning that exit conditions are satisfiable considering previous passes. In comparison with $\Phi^2_j$, which subsumes most of the intersection, the resulting time-state set $\Phi^4_j$ is now a narrow corridor as shown in Fig.~\ref{fig:pass4}. Notably, a large $\Delta$ means a large exit time window, which in turn leads to a larger $\Phi^4_j$. Yet, if $\Delta$ is too small, there might not be any feasible trajectory that satisfy all conditions. Therefore, $\Delta$ directly affects the efficiency of the intersection, and constitutes a part of the trade-off with safety.

\subsection{Driving Limits Service}

The result of \textit{Pass 4} is, for vehicle $j$, a corridor that is safe to traverse. With this in mind, we suggest the design of a service that issues driving limits such that the vehicle is ensured to stay inside its safe corridor.

For a set of states, our service computes least-restrictive control sets $A_j(z_j, t)$ that guarantee satisfaction of~\eqref{eq:ltl}. To compute the driving limits, we follow the steps described in \cite[Section V.B]{jiang_guaranteed_2024}. Let $V_{\Phi^4_j}(z_j, t)$ be the underlying value function in the HJ reachability analysis representing the safe corridor. Using $V_{\Phi^4_j}$, we compute $A_j(z_j, t)$ as the half-space $a + b^\top u_j \le 0$ with
\[\begin{split}
    a = \;  & V_{\Phi^4_j}(z_j, t) + D_t V_{\Phi^4_j}(z_j, t) \delta t \\
            & + D_z V_{\Phi^4_j}(z_j, t)^\top f_j(z_j, u_j) \delta t \\[1ex]
    b^\top = \;  & D_z V_{\Phi^4_j}(z_j, t)^\top g_j(z_j) \delta t.
\end{split}\]
Taken over time, this is a set of admissible accelerations such as depicted in Fig.~\ref{fig:ltms}. For further details, we refer to \cite{jiang_guaranteed_2024}.

Using the developed verification method in \cite{munhoz_ensuring_2024}, we can define safety specifications with the expressiveness of temporal logic. In this work, we extend this framework to incorporate entry- and exit-conditions that help address the problem of permissible planning, and we propose a service that issues driving limits that ensure vehicles remain in their reserved corridor. In the following section, we investigate if this is practically feasible with current compute capabilities.

\begin{figure*}[t]
    \centering
    \begin{subfigure}{0.3\textwidth}
        \centering
        \includegraphics[width=\textwidth]{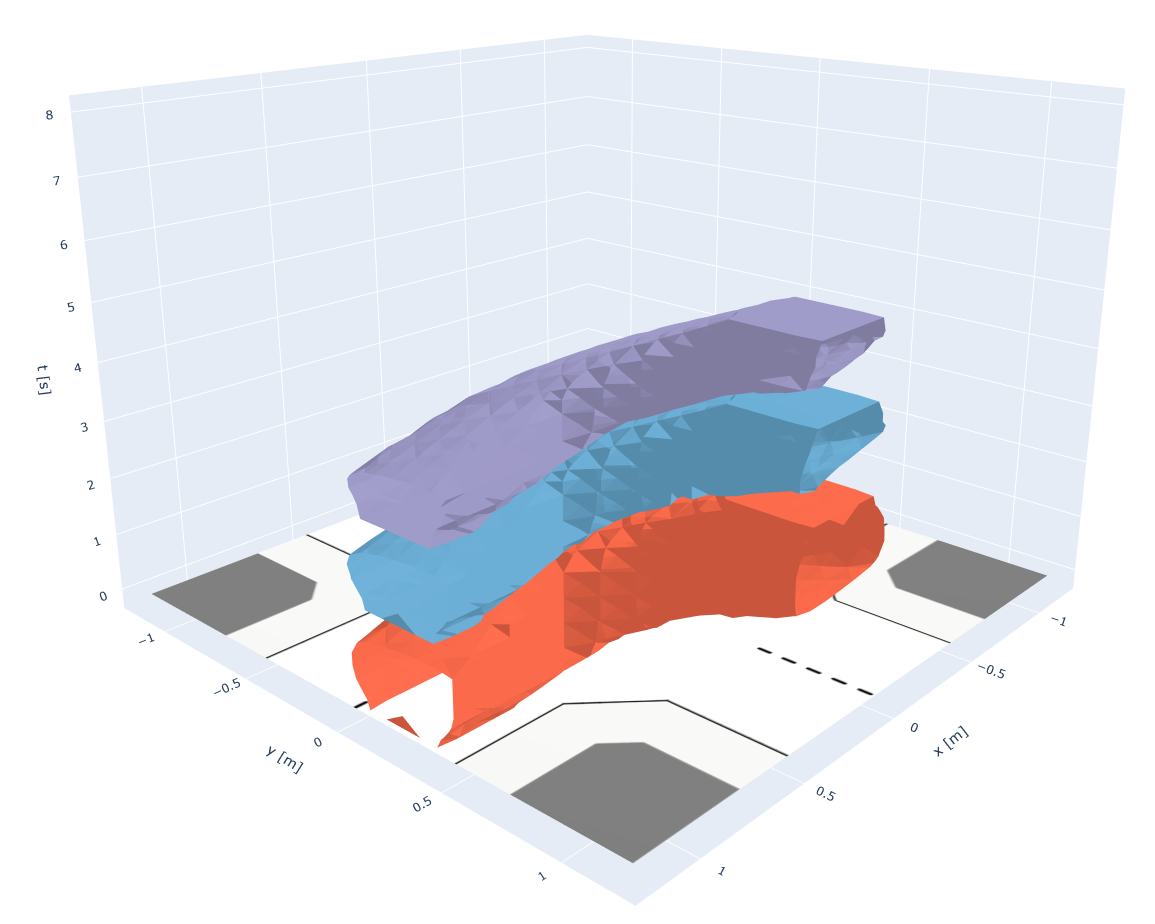}
        \caption{Three consecutive right turns.}
        \label{fig:res1}
    \end{subfigure}
    \hfill
    \begin{subfigure}{0.3\textwidth}
        \centering
        \includegraphics[width=\textwidth]{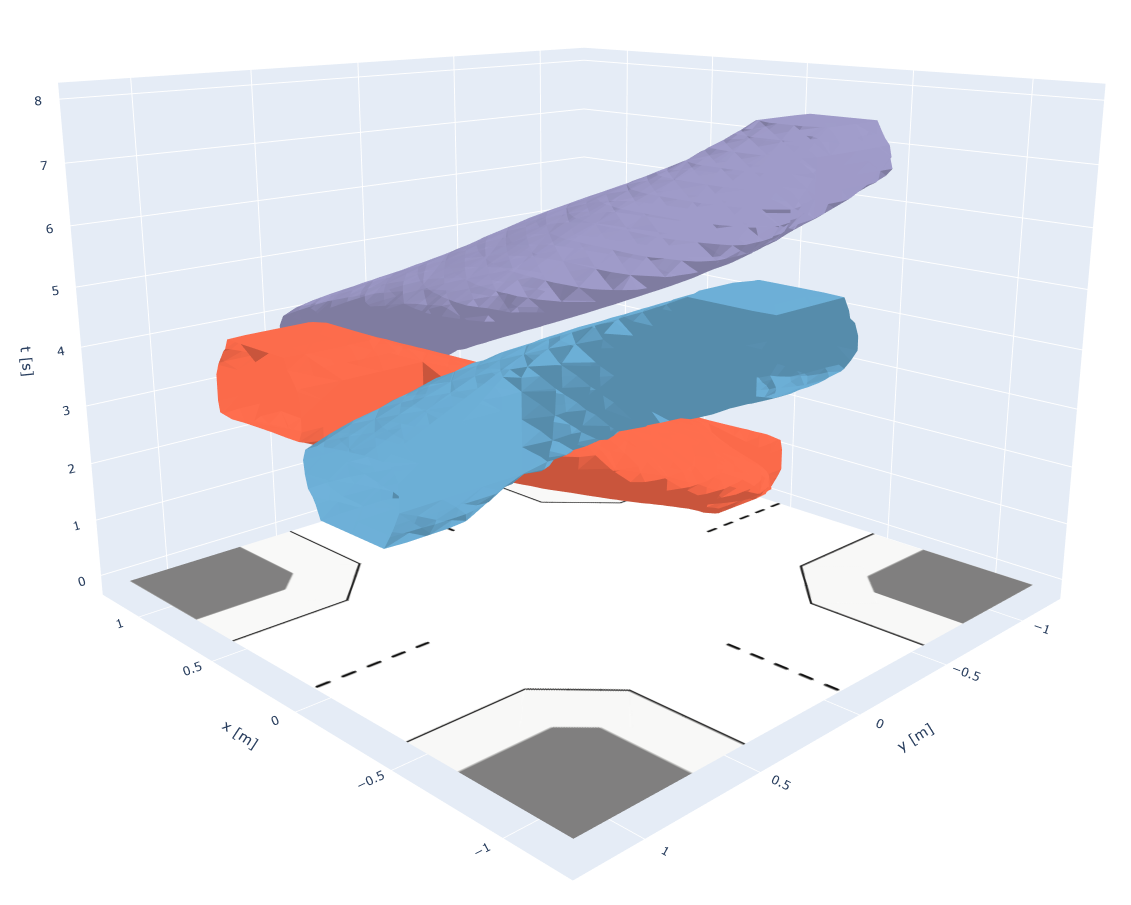}
        \caption{Straight, right, and a left turn.}
        \label{fig:res2}
    \end{subfigure}
    \hfill
    \begin{subfigure}{0.3\textwidth}
        \centering
        \includegraphics[width=\textwidth]{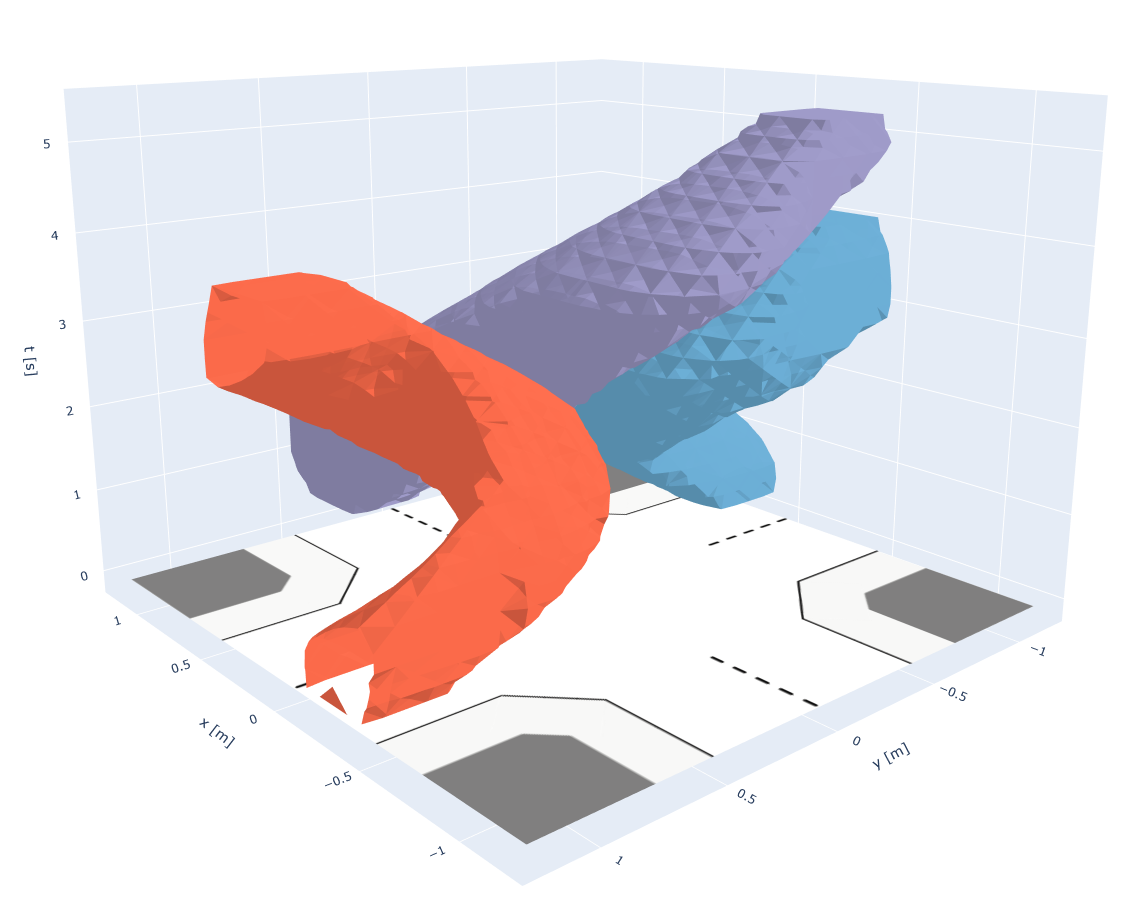}
        \caption{Two U-turns and a left turn.}
        \label{fig:res3}
    \end{subfigure}
    \caption{Shown are scenarios that illustrate the flexibility of our approach. By providing only $\mathcal G^\leftarrow_j$ and $\mathbb G^\rightarrow_j$, it is possible to compute these safe corridors.}
    \label{fig:result}
\end{figure*}

\section{Numerical Results}\label{sec:exp}

To demonstrate this work, we present numerical results that highlight the benefits of our method. First, we show that the illustrations in Fig.~\ref{fig:passes} reflect the computations of \textit{Pass~3} and \textit{Pass~4}. Next, we use the computed time-state sets to calculate least-restrictive control sets that runs the driving limits service. Finally, we end this section with examples of three-vehicle scenarios and report their computational time. The code for all results is publicly available on GitHub\footnote{\url{https://github.com/kaarmu/safe_intersections}}.

\subsection{Perpendicular Two-Vehicle Scenario}

In this scenario, two vehicles are at the 4-way intersection shown in Fig.~\ref{fig:perp-xy}. We are interested in following the second vehicle (blue) going straight through the intersection, left to right. It is preceded by another vehicle (red), that enters the area 0.9 s earlier, also going straight, but from bottom to top. Using entry and exit locations defined for left, right, bottom, and top, we perform the online safety verification with \textit{Pass~2}, \textit{Pass~3} and \textit{Pass~4}. In Fig.~\ref{fig:perp-xt}, we show a slice of the intersection at $y = \text{-0.3 m}$ to compare with the illustrations in Fig.~\ref{fig:passes}. Specifically, the red region is a slice of the reserved time-state set for the higher-priority vehicle, the gray region shows the second vehicle's time-state set from the safety analysis in \textit{Pass~2}, the green regions show the entry and exit conditions, and the blue region is \textit{Pass~4}'s final time-state set that is then reserved for the second vehicle.

\subsection{Output of Driving Limits Service}

In this work, we have suggested that vehicles at the intersection subscribe to a driving limits service. For a given set of states, the service returns saturation points for the vehicle's control inputs that ensure it stays within its reserved time-state set. For example, at $x = -0.50, y = -0.25$ in Fig.~\ref{fig:perp-xy}, the second vehicle (blue) must adhere to control constraints given by Fig.~\ref{fig:lrcs}. That is, when $0.2 \le v < 0.6$, the vehicle can safely utilize the full acceleration space. However, if standing still, the vehicle is not allowed to decelerate. Naturally, we can also see that there is no allowed control for $v \ge 0.6$ since this is above the region's speed limit and thus already outside the safe time-state set.

\subsection{Multiple Three-Vehicle Scenarios}

Lastly, we show three vehicles approaching the 4-way intersection in three different scenarios. For each scenario, the vehicles take different routes starting at slightly different times. Before reaching the intersection, each vehicle $j \in \{1,2,3\}$ requests a safe corridor from $\mathcal G^\leftarrow_j$ to $\mathbb G^\rightarrow_j$, corresponding to their selected route, with $\Delta = \text{2 s}$. In Fig.~\ref{fig:result} we show the safe corridors in 
dimensions $x, y$ and $t$, 
using colors corresponding to the priority of the vehicles. In this case, priority is given according to FCFS principles, with red, blue, and purple corridors being in highest-to-lowest order.

The first scenario seen in Fig.~\ref{fig:res1} places all vehicles on the same route, taking a right-turn at the intersection. As vehicle 1 (red) is reserving a corridor, vehicle 2 (blue) also requests to pass through the region. The intelligent intersection reserves a second corridor tightly after the first one, yet never so they overlap. Finally, vehicle 3 (purple) makes the same request. Due to being placed shortly next to each other, the two latter corridors are noticeably smaller. 
The second scenario seen in Fig.~\ref{fig:res2} presents a more dynamic situation. The same vehicles are now passing through the intersection using different routes. We also see that, if determined safe, vehicles can be scheduled at the same time. In this case, vehicle 2 (blue) and vehicle 3 (purple) drive in opposite directions, making it possible to safely pass through simultaneously.
Finally, the third scenario seen in Fig.~\ref{fig:res3} shows that our method is flexible enough to accommodate non-conventional routes. At the same time, vehicles 1 (blue) and 2 (red) request to do mirrored U-turns. Shortly after, vehicle 3 (purple) wishes to drive left in the intersection, following the first vehicle.

These examples demonstrate that our method is highly flexible and can be used for various scenarios. For each scenario and vehicle, we show the computational time of the online verification in Table~\ref{tbl:comptime}. The scenarios were simulated on a system equipped with an AMD Ryzen Threadripper 2970X processor and an NVIDIA GeForce RTX 2080 Ti graphics card. Moreover, to measure the performance of the driving limits service, we compute least-restrictive control sets for 100 randomly selected time-states for each scenario and vehicle. From this, the average computational time of a driving limits request was below 2 ms in all cases. These results indicate that, with modern software tools and hardware, our method is computationally feasible considering the real-time requirements of intelligent intersections.

\vfill

\begin{table}[t]
\centering
\renewcommand{\arraystretch}{1.5}
\begin{tabular}{cc|cccc}
           & 
           & \textbf{\renewcommand{\arraystretch}{1}\begin{tabular}[c]{@{}c@{}}Pass 2 {[s]}\end{tabular}}
           & \textbf{\renewcommand{\arraystretch}{1}\begin{tabular}[c]{@{}c@{}}Pass 3 {[s]}\end{tabular}}
           & \textbf{\renewcommand{\arraystretch}{1}\begin{tabular}[c]{@{}c@{}}Pass 4 {[s]}\end{tabular}}
           & \textbf{\renewcommand{\arraystretch}{1}\begin{tabular}[c]{@{}c@{}}Total {[s]}\end{tabular}}
           \\ 
\hline
\multirow{3}{*}{\rotatebox[origin=c]{90}{ Scenario 1 }} & Vehicle 1 & 0    & 1.39 & 1.14 & \textbf{2.57} \\
                                                        & Vehicle 2 & 0.86 & 0.17 & 0.89 & \textbf{1.98} \\
                                                        & Vehicle 3 & 0    & 0.18 & 0.93 & \textbf{1.16} \\
\hline
\multirow{3}{*}{\rotatebox[origin=c]{90}{ Scenario 2 }} & Vehicle 1 & 0    & 1.36 & 1.23 & \textbf{2.63} \\
                                                        & Vehicle 2 & 0.88 & 0.18 & 0.90 & \textbf{2.02} \\
                                                        & Vehicle 3 & 0.90 & 0.18 & 0.95 & \textbf{2.10} \\
\hline
\multirow{3}{*}{\rotatebox[origin=c]{90}{ Scenario 3 }} & Vehicle 1 & 0    & 1.33 & 1.11 & \textbf{2.48} \\
                                                        & Vehicle 2 & 0.84 & 0.17 & 0.12 & \textbf{1.20} \\
                                                        & Vehicle 3 & 0.12 & 0.16 & 0.95 & \textbf{1.29} \\
\hline
\end{tabular}
\caption{Computational time for online safety verification. Computed with AMD Ryzen Threadripper 2970X and NVIDIA GeForce RTX 2080 Ti.}
\label{tbl:comptime}
\end{table}

\section{Conclusion}\label{sec:conc}

In this paper, we develop a method to improve AIM safety by computing safe time-state corridors using reachability analysis. By addressing the problem of permissible planning, we can explicitly manage the trade-off between safety and efficiency. Additionally, we propose the driving limits service as a safety filter for AIM. Specifically, a vehicle can request a safe corridor before entering the intersection. Once inside, the vehicle can use the service to compute input saturation points, ensuring it remains within the safe corridor. To demonstrate our method, we show multiple scenarios of a simulated 4-way intersection. The results show promising performance, indicating feasibility for intelligent intersections and opportunities for integration with AIM.

In future work we wish to further improve performance using techniques such as \cite{he2023efficient}. We also plan to implement an intelligent intersection for small-scale connected and automated vehicles. Through this, we aim to evaluate the performance of our framework with real hardware and communication similar to \cite{MunhozArfvidsson2024}. Finally, an important future work will be to study how to compose safety analyses for inter-regional coordination.

\balance

\bibliographystyle{IEEEtran}
\bibliography{references,zotero}

\begin{thebibliography}{10}
\providecommand{\url}[1]{#1}
\csname url@samestyle\endcsname
\providecommand{\newblock}{\relax}
\providecommand{\bibinfo}[2]{#2}
\providecommand{\BIBentrySTDinterwordspacing}{\spaceskip=0pt\relax}
\providecommand{\BIBentryALTinterwordstretchfactor}{4}
\providecommand{\BIBentryALTinterwordspacing}{\spaceskip=\fontdimen2\font plus
\BIBentryALTinterwordstretchfactor\fontdimen3\font minus \fontdimen4\font\relax}
\providecommand{\BIBforeignlanguage}[2]{{%
\expandafter\ifx\csname l@#1\endcsname\relax
\typeout{** WARNING: IEEEtran.bst: No hyphenation pattern has been}%
\typeout{** loaded for the language `#1'. Using the pattern for}%
\typeout{** the default language instead.}%
\else
\language=\csname l@#1\endcsname
\fi
#2}}
\providecommand{\BIBdecl}{\relax}
\BIBdecl

\bibitem{arthurs_taxonomy_2022}
\BIBentryALTinterwordspacing
P.~Arthurs, L.~Gillam, P.~Krause, N.~Wang, K.~Halder, and A.~Mouzakitis, ``\BIBforeignlanguage{en}{A {Taxonomy} and {Survey} of {Edge} {Cloud} {Computing} for {Intelligent} {Transportation} {Systems} and {Connected} {Vehicles}},'' \emph{\BIBforeignlanguage{en}{IEEE Transactions on Intelligent Transportation Systems}}, vol.~23, no.~7, pp. 6206--6221, Jul. 2022. [Online]. Available: \url{https://ieeexplore.ieee.org/document/9447825/}
\BIBentrySTDinterwordspacing

\bibitem{soto_survey_2022}
\BIBentryALTinterwordspacing
I.~Soto, M.~Calderon, O.~Amador, and M.~Urueña, ``\BIBforeignlanguage{en}{A survey on road safety and traffic efficiency vehicular applications based on {C}-{V2X} technologies},'' \emph{\BIBforeignlanguage{en}{Vehicular Communications}}, vol.~33, p. 100428, Jan. 2022. [Online]. Available: \url{https://linkinghub.elsevier.com/retrieve/pii/S2214209621000978}
\BIBentrySTDinterwordspacing

\bibitem{gong_edge_2023}
\BIBentryALTinterwordspacing
T.~Gong, L.~Zhu, F.~R. Yu, and T.~Tang, ``\BIBforeignlanguage{en}{Edge {Intelligence} in {Intelligent} {Transportation} {Systems}: {A} {Survey}},'' \emph{\BIBforeignlanguage{en}{IEEE Transactions on Intelligent Transportation Systems}}, vol.~24, no.~9, pp. 8919--8944, Sep. 2023. [Online]. Available: \url{https://ieeexplore.ieee.org/document/10133894/}
\BIBentrySTDinterwordspacing

\bibitem{chen_cooperative_2016}
\BIBentryALTinterwordspacing
L.~Chen and C.~Englund, ``\BIBforeignlanguage{en}{Cooperative {Intersection} {Management}: {A} {Survey}},'' \emph{\BIBforeignlanguage{en}{IEEE Transactions on Intelligent Transportation Systems}}, vol.~17, no.~2, pp. 570--586, Feb. 2016. [Online]. Available: \url{http://ieeexplore.ieee.org/document/7244203/}
\BIBentrySTDinterwordspacing

\bibitem{ahn_semi-autonomous_2016}
\BIBentryALTinterwordspacing
H.~Ahn and D.~Del~Vecchio, ``\BIBforeignlanguage{en}{Semi-autonomous {Intersection} {Collision} {Avoidance} through {Job}-shop {Scheduling}},'' in \emph{\BIBforeignlanguage{en}{Proceedings of the 19th {International} {Conference} on {Hybrid} {Systems}: {Computation} and {Control}}}.\hskip 1em plus 0.5em minus 0.4em\relax Vienna Austria: ACM, Apr. 2016, pp. 185--194. [Online]. Available: \url{https://dl.acm.org/doi/10.1145/2883817.2883830}
\BIBentrySTDinterwordspacing

\bibitem{altche_least_2016}
\BIBentryALTinterwordspacing
F.~Altche, X.~Qian, and A.~De~La~Fortelle, ``\BIBforeignlanguage{en}{Least restrictive and minimally deviating supervisor for {Safe} semi-autonomous driving at an intersection: {An} {MIQP} approach},'' in \emph{\BIBforeignlanguage{en}{2016 {IEEE} 19th {International} {Conference} on {Intelligent} {Transportation} {Systems} ({ITSC})}}.\hskip 1em plus 0.5em minus 0.4em\relax Rio de Janeiro, Brazil: IEEE, Nov. 2016, pp. 2520--2526. [Online]. Available: \url{http://ieeexplore.ieee.org/document/7795961/}
\BIBentrySTDinterwordspacing

\bibitem{dresner_multiagent_2008}
\BIBentryALTinterwordspacing
K.~Dresner and P.~Stone, ``\BIBforeignlanguage{en}{A {Multiagent} {Approach} to {Autonomous} {Intersection} {Management}},'' \emph{\BIBforeignlanguage{en}{Journal of Artificial Intelligence Research}}, vol.~31, pp. 591--656, Mar. 2008. [Online]. Available: \url{https://jair.org/index.php/jair/article/view/10542}
\BIBentrySTDinterwordspacing

\bibitem{chamideh_safe_2023}
\BIBentryALTinterwordspacing
S.~Chamideh, W.~Tärneberg, and M.~Kihl, ``\BIBforeignlanguage{en}{A {Safe} and {Robust} {Autonomous} {Intersection} {Management} {System} {Using} a {Hierarchical} {Control} {Strategy} and {V2I} {Communication}},'' \emph{\BIBforeignlanguage{en}{IEEE Systems Journal}}, vol.~17, no.~1, pp. 50--61, Mar. 2023. [Online]. Available: \url{https://ieeexplore.ieee.org/document/9962817/}
\BIBentrySTDinterwordspacing

\bibitem{namazi_intelligent_2019}
\BIBentryALTinterwordspacing
E.~Namazi, J.~Li, and C.~Lu, ``\BIBforeignlanguage{en}{Intelligent {Intersection} {Management} {Systems} {Considering} {Autonomous} {Vehicles}: {A} {Systematic} {Literature} {Review}},'' \emph{\BIBforeignlanguage{en}{IEEE Access}}, vol.~7, pp. 91\,946--91\,965, 2019. [Online]. Available: \url{https://ieeexplore.ieee.org/document/8756239/}
\BIBentrySTDinterwordspacing

\bibitem{colombo_efficient_2012}
\BIBentryALTinterwordspacing
A.~Colombo and D.~Del~Vecchio, ``\BIBforeignlanguage{en}{Efficient algorithms for collision avoidance at intersections},'' in \emph{\BIBforeignlanguage{en}{Proceedings of the 15th {ACM} international conference on {Hybrid} {Systems}: {Computation} and {Control}}}.\hskip 1em plus 0.5em minus 0.4em\relax Beijing China: ACM, Apr. 2012, pp. 145--154. [Online]. Available: \url{https://dl.acm.org/doi/10.1145/2185632.2185656}
\BIBentrySTDinterwordspacing

\bibitem{hafner_cooperative_2013}
\BIBentryALTinterwordspacing
M.~R. Hafner, D.~Cunningham, L.~Caminiti, and D.~Del~Vecchio, ``\BIBforeignlanguage{en}{Cooperative {Collision} {Avoidance} at {Intersections}: {Algorithms} and {Experiments}},'' \emph{\BIBforeignlanguage{en}{IEEE Transactions on Intelligent Transportation Systems}}, vol.~14, no.~3, pp. 1162--1175, Sep. 2013. [Online]. Available: \url{https://ieeexplore.ieee.org/document/6495719}
\BIBentrySTDinterwordspacing

\bibitem{saraoglu_designing_2022}
\BIBentryALTinterwordspacing
M.~Saraoglu, J.~Pintscher, and K.~Janschek, ``\BIBforeignlanguage{en}{Designing a {Safe} {Intersection} {Management} {Algorithm} using {Formal} {Methods}},'' \emph{\BIBforeignlanguage{en}{IFAC-PapersOnLine}}, vol.~55, no.~14, pp. 22--27, 2022. [Online]. Available: \url{https://linkinghub.elsevier.com/retrieve/pii/S2405896322009880}
\BIBentrySTDinterwordspacing

\bibitem{irani_liu_specification-compliant_2023}
\BIBentryALTinterwordspacing
E.~Irani~Liu and M.~Althoff, ``\BIBforeignlanguage{en}{Specification-{Compliant} {Driving} {Corridors} for {Motion} {Planning} of {Automated} {Vehicles}},'' \emph{\BIBforeignlanguage{en}{IEEE Transactions on Intelligent Vehicles}}, vol.~8, no.~9, pp. 4180--4197, Sep. 2023. [Online]. Available: \url{https://ieeexplore.ieee.org/document/10163843/}
\BIBentrySTDinterwordspacing

\bibitem{munhoz_ensuring_2024}
\BIBentryALTinterwordspacing
K.~M. Arfvidsson, F.~J. Jiang, K.~H. Johansson, and J.~Mårtensson, ``{Ensuring Safety at Intelligent Intersections: Temporal Logic Meets Reachability Analysis},'' 2024. [Online]. Available: \url{https://arxiv.org/abs/2405.11300}
\BIBentrySTDinterwordspacing

\bibitem{jiang_guaranteed_2024}
\BIBentryALTinterwordspacing
F.~J. Jiang, K.~M. Arfvidsson, C.~He, M.~Chen, and K.~H. Johansson, ``\BIBforeignlanguage{en}{Guaranteed {Completion} of {Complex} {Tasks} via {Temporal} {Logic} {Trees} and {Hamilton}-{Jacobi} {Reachability}},'' Apr. 2024, arXiv:2404.08334 [cs, eess]. [Online]. Available: \url{http://arxiv.org/abs/2404.08334}
\BIBentrySTDinterwordspacing

\bibitem{maierhofer_formalization_2020}
\BIBentryALTinterwordspacing
S.~Maierhofer, A.-K. Rettinger, E.~C. Mayer, and M.~Althoff, ``\BIBforeignlanguage{en}{Formalization of {Interstate} {Traffic} {Rules} in {Temporal} {Logic}},'' in \emph{\BIBforeignlanguage{en}{2020 {IEEE} {Intelligent} {Vehicles} {Symposium} ({IV})}}.\hskip 1em plus 0.5em minus 0.4em\relax Las Vegas, NV, USA: IEEE, Oct. 2020, pp. 752--759. [Online]. Available: \url{https://ieeexplore.ieee.org/document/9304549/}
\BIBentrySTDinterwordspacing

\bibitem{bansal_provably_2021}
S.~Bansal, M.~Chen, K.~Tanabe, and C.~J. Tomlin, ``Provably {Safe} and {Scalable} {Multivehicle} {Trajectory} {Planning},'' \emph{IEEE Transactions on Control Systems Technology}, vol.~29, no.~6, pp. 2473--2489, Nov. 2021, conference Name: IEEE Transactions on Control Systems Technology.

\bibitem{he2023efficient}
C.~He, Z.~Gong, M.~Chen, and S.~Herbert, ``{Efficient and Guaranteed Hamilton-Jacobi Reachability via Self-Contained Subsystem Decomposition and Admissible Control Sets},'' \emph{IEEE Control Systems Letters}, 2023.

\bibitem{MunhozArfvidsson2024}
\BIBentryALTinterwordspacing
K.~M. Arfvidsson, K.~Fragkedaki, F.~J. Jiang, V.~Narri, H.-C. Lindh, K.~H. Johansson, and J.~Mårtensson, ``{Small-Scale Testbed for Evaluating C-V2X Applications on 5G Cellular Networks},'' 2024. [Online]. Available: \url{https://arxiv.org/abs/2405.05911}
\BIBentrySTDinterwordspacing

\end{thebibliography}

\end{document}